\documentclass{jfm}

\usepackage{graphicx}
\usepackage{newtxtext}
\usepackage{newtxmath}
\usepackage{natbib}
\usepackage{hyperref}
\usepackage{multirow}
\usepackage{amsfonts}
\usepackage{amssymb}
\usepackage{bm}

\hypersetup{
    colorlinks = true,
    urlcolor   = blue,
    citecolor  = black,
}

\newcommand{\RomanNumeralCaps}[1]
\linenumbers

\title{No sustained mean velocity in the boundary region of plane thermal convection}

\author{
Roshan J. Samuel\aff{1},
Mathis Bode\aff{2},
Janet D. Scheel \aff{3}
Katepalli R. Sreenivasan\aff{4,5,6}
\and J\"org Schumacher\aff{1,4} \corresp{\email{joerg.schumacher@tu-ilmenau.de}}
}

\affiliation{
\aff{1}Institute of Thermodynamics and Fluid Mechanics, Technische Universit\"{a}t Ilmenau, D-98684 Ilmenau, Germany
\aff{2}J\"{u}lich Supercomputing Centre, Forschungszentrum Jülich GmbH, D-52428 J\"{u}lich, Germany
\aff{3}Department of Physics, Occidental College, Los Angeles, CA 90041, USA
\aff{4}Tandon School of Engineering, New York University, New York, NY 11201, USA
\aff{5}Courant Institute of Mathematical Sciences, New York University, New York, NY 10012, USA
\aff{6}Center for Space Science, New York University, Abu Dhabi, 100565, UAE
}

\begin{document}
\maketitle

\begin{abstract}
We study the dynamics of thermal and momentum boundary regions in three-dimensional direct numerical simulations of Rayleigh-B\'{e}nard convection for the Rayleigh number range $10^5\le Ra \le 10^{11}$ and $Pr=0.7$.
Using a Cartesian slab with horizontal periodic boundary conditions and an aspect ratio of 4, we obtain statistical homogeneity in the horizontal $x$- and $y$-directions, thus approximating best an extended convection layer relevant for most geo- and astrophysical flow applications. 
We observe upon canonical use of combined long-time and area averages, with averaging periods of at least 100 free-fall times, that a global coherent mean flow is practically absent and that the magnitude of the velocity fluctuations is larger than the mean by up to 2 orders of magnitude.
The velocity field close to the wall is a collection of differently oriented local shear-dominated flow patches interspersed by extensive shear-free incoherent regions which can be as large as the whole cross section, unlike for a closed cylindrical convection cell of aspect ratio of the order 1.
The incoherent regions occupy a 60\% area fraction for all Rayleigh numbers investigated here.
Rather than resulting in a pronounced mean flow with small fluctuations about such a mean, as found in small-aspect-ratio convection, the velocity field is dominated by strong fluctuations of all three components around a non-existent or weak mean. We discuss the consequences of these observations for convection layers with larger aspect ratios, including boundary layer instabilities and the resulting turbulent heat transport.
\end{abstract}

\begin{keywords}
Rayleigh-B\'{e}nard convection, boundary layers
\end{keywords}

\section{Introduction}
\label{sec:intro}
Rayleigh-B\'enard convection (RBC) is one of the fundamental flow configurations in fluid turbulence research. The fluid in this configuration is nominally confined to an infinitely extended layer enclosed between two parallel, horizontal and impermeable plates separated by a vertical distance $H$ \citep{Rayleigh1916}. When the fluid layer is heated sufficiently strongly from below (and cooled from above), buoyancy forces initiate a turbulent fluid motion that has a statistically preferred state with respect to the direction of gravity, ${\bm g} = {g \bm e}_z$. A central question concerns the amount and nature of heat and momentum carried through the layer, and their dependencies on the imposed temperature difference between the top and bottom plates, $\Delta T= T_{\rm bot}-T_{\rm top}$. The temperature difference is expressed by the dimensionless Rayleigh number $Ra=g\alpha \Delta T H^3/(\nu \kappa)$, where $\alpha$ is the isobaric expansion coefficient, $\nu$ the kinematic viscosity, and $\kappa$ the temperature diffusivity \citep{Ahlers:RMP2009,Chilla:EPJE2012,Verma:BDF}. 

A second important question is the structure of the velocity and thermal boundary layers on the horizontal walls and their effects on heat transport. Since the RBC system is enclosed by walls at the top and bottom, the viscous and thermal boundary layers formed on these walls pose a bottleneck for the global transport of both heat and momentum. Their composition and dynamics at very high Rayleigh numbers still need to be better understood as emphasized recently \citep{Iyer:PNAS2020,Lindborg2023,Shishkina2023,Creyssels2024}. For Rayleigh numbers $Ra\gtrsim 10^{10}$, no laboratory experiment to-date has resolved the {\em dynamic} interplay of the boundary layers and their fluctuations with the basic structural elements, namely thermal plumes and shear layers. Direct numerical simulations (DNS) are thus the only way to compare their structure and statistical properties with predictions from theories for canonical laminar and turbulent boundary layers \citep{Schlichting:book:BL}. Furthermore, the closed-cell geometry of high-Rayleigh-number studies \citep{Castaing:JFM1989,Chavanne:PRL1997,Niemela:Nature2000,Ahlers:JFM2009RBC,Urban:PRL2012} breaks the horizontal translation symmetry of the statistics, except possibly when the aspect ratio is very large \citep{Pandey:NC2018}. Small aspect ratio enforces a dominant large-scale circulation (LSC) in the cell \citep{Kadanoff:PT2001}, manifesting as a relatively coherent shear flow connecting the top and bottom plates, fluctuating only moderately in its mean orientation \citep{Sreenivasan:PRE2002, Stevens:JFM2011,Shi:JFM2012,Scheel:PRF2017}. Both aspects take us away from the original question on the heat and momentum transfer in an {\em infinitely extended} plane layer.

In this work, we focus more on the second question in a configuration that is closest to the original RBC model of convection between a pair of infinitely extended planes, by using periodic boundaries in both horizontal directions. Simulations with similar boundary conditions and Prandtl numbers have been done by  \cite{Kerr:JFM1996} for $\Gamma=6$ and $Ra\le 2\times 10^7$, by \cite{Hartlep:PRL2003} for $\Gamma = 10$ and $Ra\le 1\times 10^7$, by \cite{vanReeuwijk:PRE2008a} and \cite{vanReeuwijk:PRE2008b} for $\Gamma = 4$ and $Ra\le 1\times 10^8$, by \cite{De:EJMB2018} for $\Gamma = 6$ but $Ra \le 2\times 10^6$, and by \cite{Stevens:PRF2018} for $\Gamma\le 32$ at $Ra=10^9$. Our DNS spans Rayleigh numbers of six orders of magnitude up to $Ra=10^{11}$ for long periods of time (see table 1 for details). The choice of an aspect ratio of 4 for the present study provides a ``sweet spot''. On the one hand, the domain is large enough that it does not generate strong large scale circulations, see \citet{Niemela:JFM2006}. On the other hand, it is small enough to allow us to advance to very high Rayleigh numbers, here of up to $Ra=10^{11}$, since the required numerical resources grow with $\Gamma^2$. Furthermore, this is the aspect ratio beyond which Nusselt and Reynolds numbers become independent of $\Gamma$ as discussed in \citet{Stevens:PRF2018}. We supplement these results by additional DNS runs at $\Gamma=2$ and 8 for $Ra=10^9$.

A canonical mean flow analysis reveals practically no global mean flow; instead, strong velocity fluctuations dominate the flow at all $Ra$. Fits to the mean vertical velocity profiles result in very small free-stream velocities $U_{\infty}\sim 10^{-3}$ in terms of the free fall velocity $U_f = \sqrt{g\alpha \Delta T H}$, and thicknesses $\delta_{\infty}\sim 10^{-2}$ in terms of $H$, resulting in small shear Reynolds numbers $Re_{\rm shear}\lesssim 1$--10 even for the largest $Ra$. We further analyse fluctuations of the velocity components, determine the distances of maximum mean-square fluctuations from the wall, and discuss the resulting Reynolds numbers. 
\begin{figure}
\centerline{\includegraphics[width=0.99\textwidth]{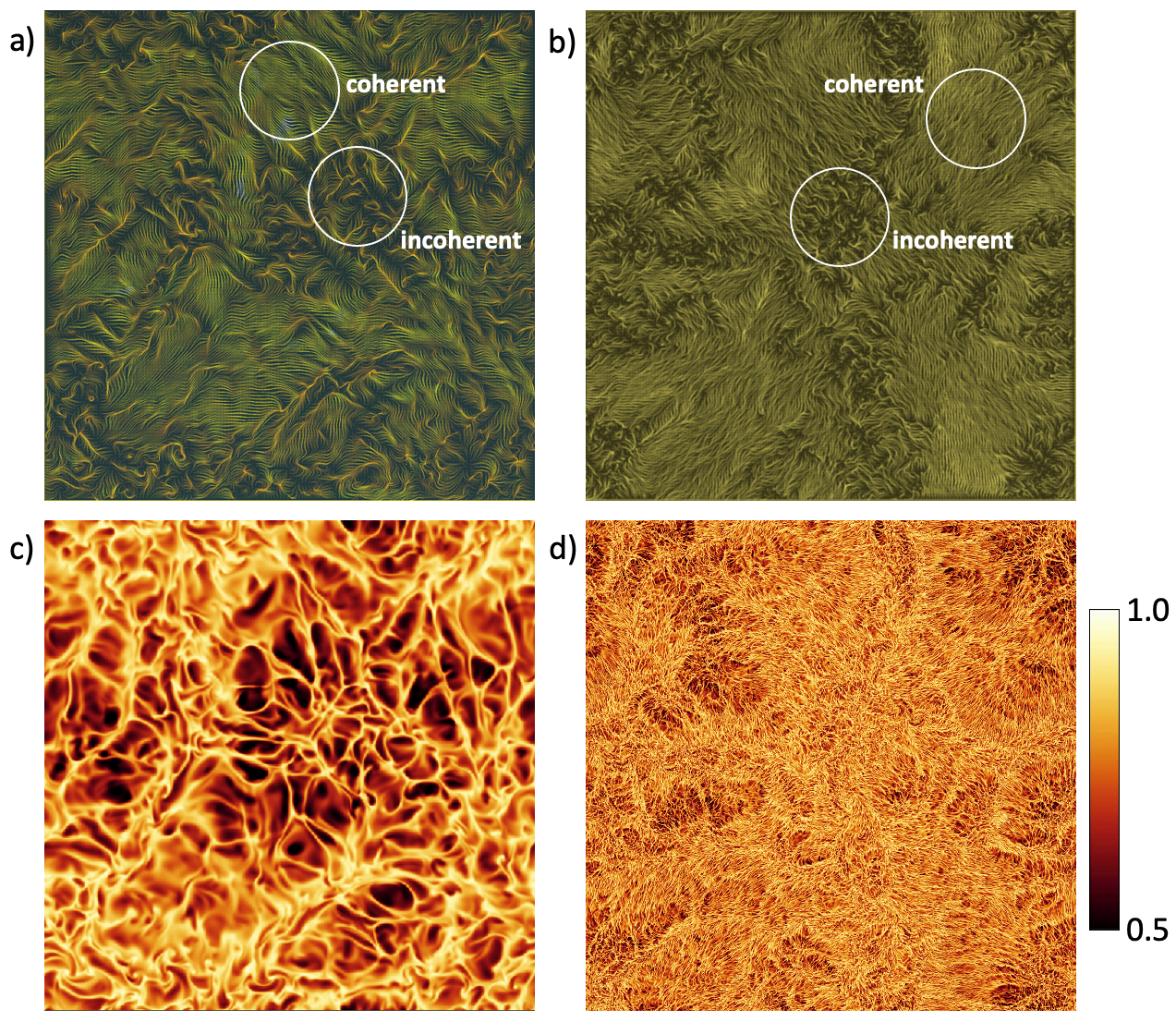}}
  \caption{Turbulent fields at two Rayleigh numbers. Streamline snapshots of the horizontal velocity field $(u_x, u_y)$ and contour plots of temperature field $T$ at the same instant. (a,c) $Ra=10^8$. (b,d) $Ra=10^{11}$. The entire cross section of $4H\times 4H$ at $z=\delta_T/2$ close to the bottom wall is shown. $H$ is the height of the convection layer and $\delta_T$ the thermal boundary layer thickness. The colour legend holds for both temperature field panels (c,d). In panels (a) and (b) examples for local coherent shear-dominated and incoherent shear-free regions are indicated.}
\label{fig:contours}
\end{figure}

A quick impression of the complex boundary layer dynamics is obtained by the streamline and contour plots in figure \ref{fig:contours} close to the bottom wall for two Rayleigh numbers. The figures indicate a prominent patchiness of the whole velocity boundary layer viewed from the top. The boundary layer is composed of coherent shear-dominated and incoherent shear-free regions. This feature becomes less prominent for the contours of the temperature field, which display an increasingly dense skeleton of thermal plume ridges over the whole plate. Here, we quantify the corresponding area fraction, condition the fluctuations on the coherent and incoherent regions, and relate the incoherent regions to the large-scale patterns in the bulk, which are the turbulent superstructures of convection \citep{Pandey:NC2018,Stevens:PRF2018}. Only when the velocity fluctuations are conditioned on the coherent shear-dominated regions are the mean profiles close to those observed in closed cylinders for $\Gamma \sim 1$ (\citet{Scheel:PRF2017}.

One important point needs to be made here. While a thermal boundary layer of the standard type is indeed present, no momentum boundary layer can be easily identified, as explained subsequently. For the velocity field, it is thus more accurate to merely discuss the flow near the wall instead of the boundary layer flow, but for convenience of identification and following convention, we continue to use the term boundary layer. There is no such ambiguity for the thermal boundary layer.

\section{Numerical simulations and resolution analysis}
\label{sec:method}

We solve the three-dimensional Boussinesq equations of RBC \citep{Verma:BDF} by the GPU-based spectral element code nekRS \citep{Fischer2022} which combines an element decomposition of the computational domain with a spectral expansion in Lagrange polynomials of each involved field along each spatial dimension on each element. The equations are given in dimensionless form by 
\begin{align}
\frac{\partial {\bm u}}{\partial t} + ({\bm u} \cdot {\bm \nabla}) {\bm u} & =  -{\bm \nabla}p + T \hat{{\bm z}} + \sqrt{\frac{Pr}{Ra}} \, \nabla^2 {\bm u}, 
\label{eq:u} \\
\frac{\partial T}{\partial t} + ({\bm u} \cdot {\bm \nabla}) T & =  \frac{1}{\sqrt{PrRa}} \, \nabla^2 T, 
\label{eq:T} \\
{\bm \nabla} \cdot {\bm u} & =  0. 
\label{eq:m}
\end{align} 
Here, ${\bm u}$, $p$, and $T$ are the velocity, pressure, and temperature fields, respectively. Length, velocity, and temperature are expressed in units of $H$, $U_f$, and the outer temperature difference $\Delta T$, respectively. No-slip boundary conditions apply for the velocity field at the plates at $z=0$ and $H$. Table \ref{tab:sim_details} summarizes 10 simulations, all at a Prandtl number $Pr=\nu/\kappa=0.7$ and aspect ratios $\Gamma=L/H=2$, 4, and 8 where $L$ is the horizontal length. The number of collocation points inside the thermal boundary layer (based on the temperature fluctuation profiles) is always $N_{BL}\ge 15$. Furthermore, we verified that the Nusselt numbers $Nu_{\rm vol}$ and $Nu_{\rm wall}$, which are given by combined volume-time $\langle\cdot\rangle_{V,t}$ and area-time averages $\langle\cdot\rangle_{A,t}$,
\begin{equation}
Nu_\mathrm{vol}=1+\sqrt{Ra Pr}\langle u_z T\rangle_{V,t} \quad\mbox{and}\quad
Nu_\mathrm{wall}=-\frac{\partial \langle T\rangle_{A,t}}{\partial z}\big|_{z=0,H}\,,
\end{equation}
result in practically the same values (table 1). 

\begin{figure}
\centerline{\includegraphics[width=0.9\textwidth]{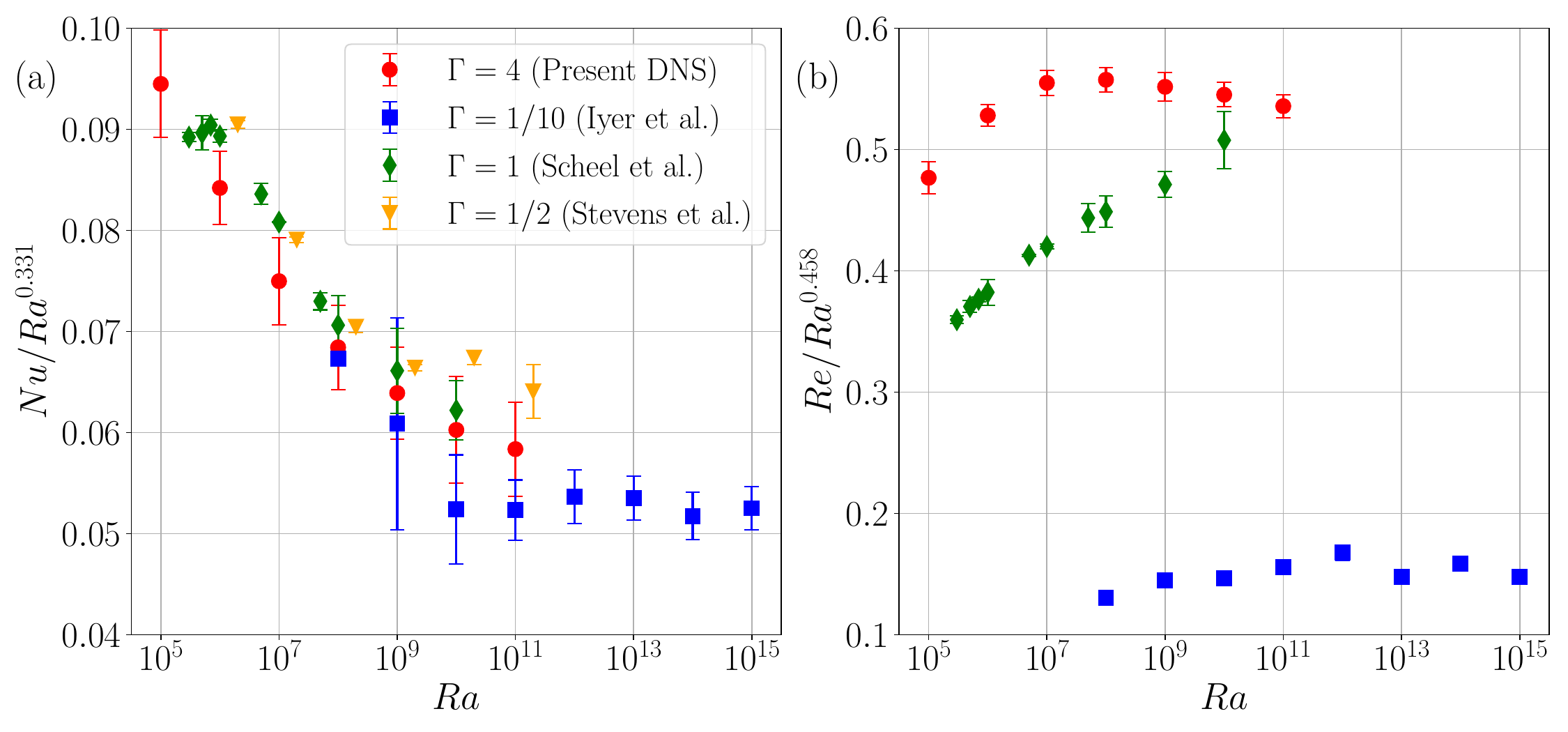}}
  \caption{(a) Compensated turbulent heat transfer $Nu/Ra^{0.331}$ versus $Ra$.
           (b) Compensated turbulent momentum transfer $Re/Ra^{0.458}$ versus $Ra$.
           The panels contain data from the present DNS and from \cite{Scheel:PRF2017}, \cite{Stevens:JFM2010} and \cite{Iyer:PNAS2020}.
           All data are compensated by the fits to the data from \cite{Iyer:PNAS2020}; see also table \ref{tab:sim_details}.} 
\label{fig:NuRe}
\end{figure}

As a first global result, we plot in figure \ref{fig:NuRe}(a,b) the Nusselt and Reynolds numbers, $Nu=Nu_{\rm vol}$ and $Re=U_{\rm rms} \sqrt{Ra/Pr}$ versus the Rayleigh number $Ra$, compensated by the high-$Ra$ scaling result of \cite{Iyer:PNAS2020}; here $U_{\rm rms}=\langle {\bm u}^2\rangle_{V,t}^{1/2}$. Also compared are our results with those of \cite{Stevens:JFM2010} and \cite{Scheel:PRF2017} in closed cylinders at $\Gamma=1/2$ and $\Gamma=1$, respectively. Considering the large differences in aspect ratios, the Nusselt numbers collapse fairly well and follow the same trend for $Ra\le 10^{11}$ (figure 2(a)); they are also in agreement with previous simulations for $Ra\le 10^8$ from \cite{vanReeuwijk:PRE2008a}. The Reynolds number shows a strong geometry dependence as visible in figure 2(b), though they tend to the same exponent towards high $Ra$; this suggests an agreement in scaling exponents at high $Ra$, but the prefactor seems to have a complicated Rayleigh number dependence. Statistics in the all runs are obtained for equal and more than 100 free-fall times $T_f=H/U_f$; see table \ref{tab:sim_details}. 

\begin{figure}
\centerline{\includegraphics[width=0.6\textwidth]{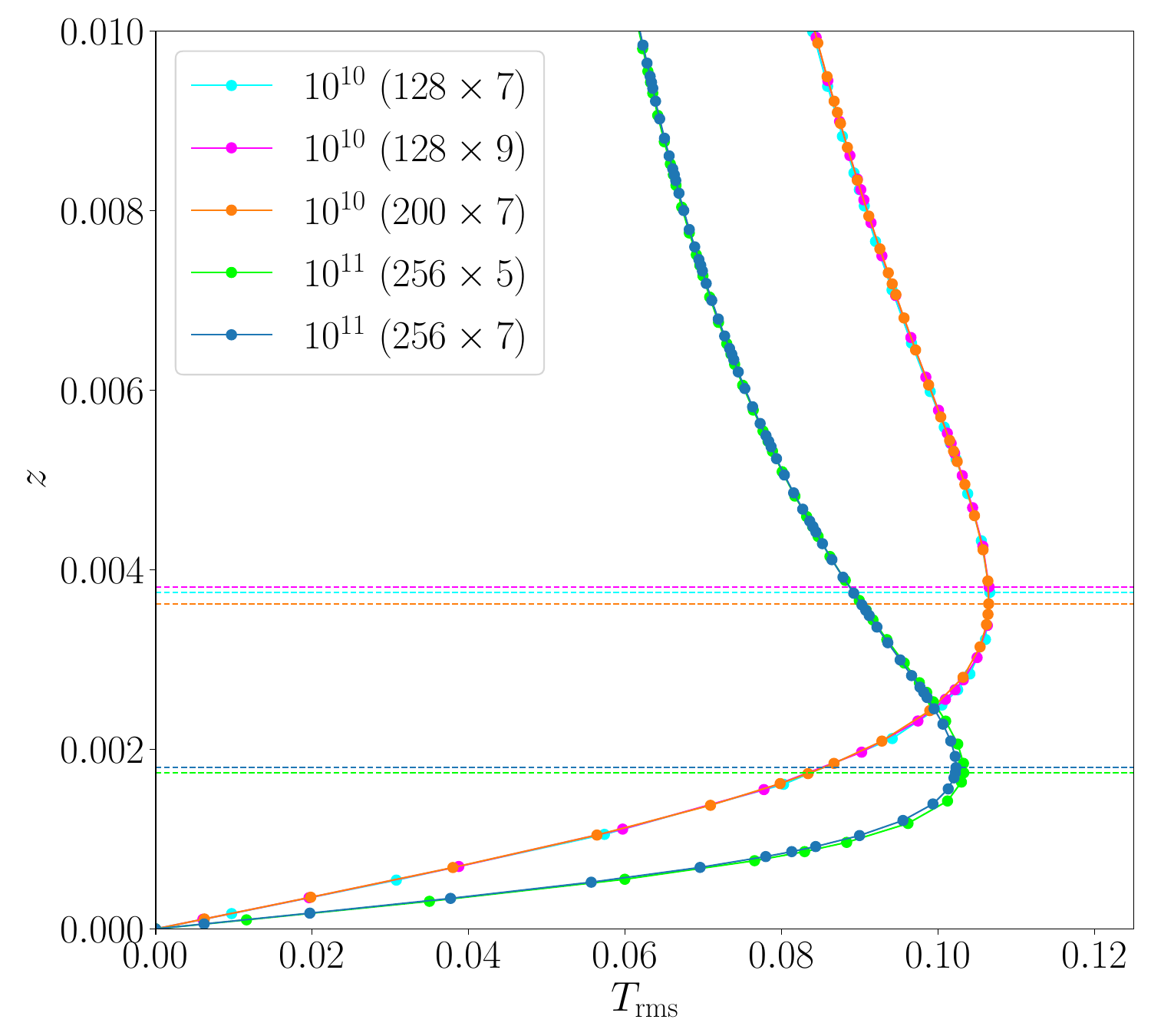}}
  \caption{Resolution study for the runs at the two highest Rayleigh numbers. The near-wall profile of the temperature fluctuations is plotted. The legend indicates the Rayleigh number together with the vertical numbers of spectral elements and the polynomial order. The runs, as listed in the legend from top to bottom, have 11, 14, 16, 11, and 16 points inside the thermal boundary layer. The horizontal lines are the thermal fluctuation boundary layer thicknesses.} 
\label{fig:resolution}
\end{figure}

We verified that the resolution of the boundary layers is sufficient. Figure \ref{fig:resolution} shows the vertical profiles of the temperature fluctuations for three different spectral element grids at $Ra=10^{10}$ and two at $Ra=10^{11}$ with different polynomial orders $p\ge 5$. It is seen that the profiles collapse well on each other, thus also demonstrating the convergence of the spectral method. This conclusion is further reinforced for $Ra=10^9$ where we have 4 DNS runs at different aspect ratio and vertical grid stretching. First, the Nusselt and Reynolds numbers in table \ref{tab:sim_details} differ only slightly (of the order of a tenth of a percent)~between $\Gamma=4$ and 8. Secondly, they are very close for two element grids with different vertical grid stretching at $\Gamma=4$. The fast spectral convergence of spectral element methods in comparison to lower-order finite-difference schemes has been reported recently by \citet{Zahtila2023} in comprehensive studies for turbulent channel flows.
\begin{table}
  \begin{center}
    \begin{tabular}{lccccccccc}
    $Ra$                & $\Gamma$ &  $N_e$         & $p$   & \(N_{BL}\) & \(\tau_{\rm total}\) &  $Nu_\mathrm{vol}$  & $Nu_\mathrm{wall}$ & $\Rey$& $Ra_{\delta,{\rm rms}}$    \\
    \(10^{5}\)          & 4     & \(100 \times 100 \times 64\)  & 5     & 71    & 1000  & $ 4.27 \pm 0.24$  & $  4.27 \pm 0.15$ & $ 93.0 \pm  2.6$  & $104 \pm 21$  \\
    \(10^{6}\)          & 4     & \(100 \times 100 \times 64\)  & 7     & 57    & 1000  & $ 8.15 \pm 0.35$  & $  8.15 \pm 0.19$ & $  296 \pm  5  $  & $116 \pm 14$  \\
    \(10^{7}\)          & 4     & \(100 \times 100 \times 64\)  & 9     & 42    & 1000  & $ 15.6 \pm 0.9 $  & $ 15.6 \pm 0.3$   & $  892 \pm  17 $  & $144 \pm 14$  \\
    \(10^{8}\)          & 4     & \(150 \times 150 \times 96\)  & 7     & 24    & 1000  & $ 30.4 \pm 1.9 $  & $ 30.4 \pm 0.4$   & $ 2571 \pm  47 $  & $174 \pm 15$  \\
    \(10^{9}\)          & 4     & \(150 \times 150 \times 96\)  & 9     & 15    & 400   & $ 60.9 \pm 4.4 $  & $ 60.8 \pm 0.9$   & $ 7300 \pm  150$  & $209 \pm 15$  \\
    \(10^{9\,\dagger}\) & 2     & \(120 \times 120 \times 96\)  & 9     & 21    & 100   & $ 62.4 \pm 5.0 $  & $ 62.4 \pm 1.2$   & $ 6990 \pm  210$  & $177 \pm 21$  \\
    \(10^{9\,\dagger}\) & 4     & \(220 \times 220 \times 96\)  & 9     & 22    & 100   & $ 60.5 \pm 3.8 $  & $ 60.5 \pm 0.7$   & $ 7340 \pm  140$  & $206 \pm 18$  \\
    \(10^{9\,\dagger}\) & 8     & \(450 \times 450 \times 96\)  & 9     & 22    & 100   & $ 60.9 \pm 2.0 $  & $ 60.9 \pm 0.3$   & $ 7180 \pm   70$  & $212 \pm 10$  \\
    \(10^{10}\)         & 4     & \(400 \times 400 \times 200\) & 7     & 16    & 100   & $122.9 \pm 10.8$  & $123.1 \pm 1.4$   & $20720 \pm  380$  & $247 \pm 22$  \\
    \(10^{11}\)         & 4     & \(500 \times 500 \times 256\) & 7     & 16    & 100   & $254 \pm 21$      & $254 \pm 2$       & $58550 \pm 1040$  & $298 \pm 23$  \\
    \end{tabular}
    \caption{
    Details of the simulation series.
    Listed here are the Rayleigh number $Ra$, the aspect ratio $\Gamma$, the number of spectral elements $N_e$,
    the polynomial order $p$ on each element with respect to each space direction,
    the number of collocation points within the thermal boundary layer $N_{BL}$,
    the total averaging time in free-fall units $\tau_{\rm total}$,
    the volume averaged and wall-averaged Nusselt numbers,
    $Nu_{\rm vol} (= Nu)$ and $Nu_\mathrm{wall}$, the Reynolds number $\Rey$,
    and the Rayleigh number based on the thermal fluctuation boundary layer thickness, $Ra_{\delta,{\rm rms}}$.
    The three DNS runs with dagger symbols at $Ra=10^9$ have a different vertical spectral element grid stretching than the fourth DNS run at this Rayleigh number.
    The total number of mesh cells is $N_e\times p^3$. It increases to almost $2.2\times 10^{10}$ collocation mesh cells for $Ra=10^{11}$.
    Mean values in the last four columns are accompanied by the standard deviations. The Prandtl number of all runs is $Pr=0.7$.}
    \label{tab:sim_details}
  \end{center}
\end{table}
\begin{figure}
\centerline{\includegraphics[width=0.95\textwidth]{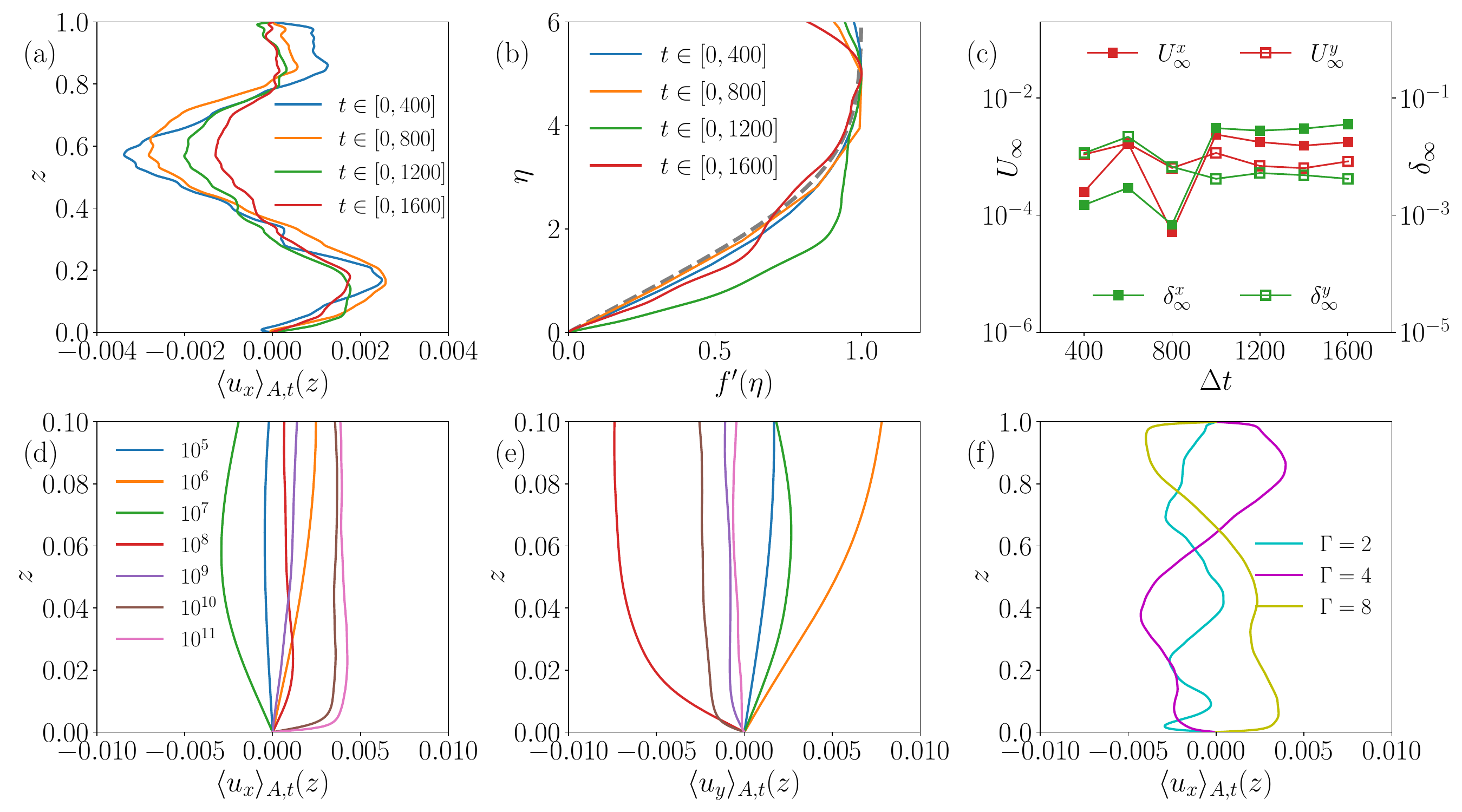}}
  \caption{(a) Planar-averaged profiles of the horizontal velocity component $u_x$ for different averaging intervals for $Ra = 10^9$. Notice the very small values of $U_{\infty}$ and the non-monotonic convergence of the profiles. (b) Match of the near-wall profiles from panel (a) to the Blasius profile $f^{\prime}(\eta)$ (dashed line). (c) Variation of the free-stream velocity $U_{\infty}$ and boundary layer height $\delta_{\infty}$ obtained from the fits to the Blasius profile versus averaging time at $Ra=10^9$. (d) Time-averaged profiles $\langle u_x\rangle_{A,t}$ for all 7 Rayleigh numbers. (e) Time-averaged profiles $\langle u_y\rangle_{A,t}$ for all 7 Rayleigh numbers. (f) Time-averaged profiles $\langle u_x\rangle_{A,t}$ for Rayleigh number $Ra=10^9$ and aspect ratios $\Gamma=2$, 4 and 8. These are the three runs which are indicated by a dagger in table \ref{tab:sim_details}.}
\label{fig:long_run_mean}
\end{figure}

\section{Mean profiles of temperature and velocity}
The mean velocity profiles for the horizontal components are obtained by a combined average over the area $A=L^2$ and $N_t=20$ statistically independent realizations of the turbulent flow separated from each other by at least 5$T_f$ as  
\begin{equation}
\langle u_i\rangle_{A,t}(z)=\frac{1}{A N_t}\sum_{k=1}^{N_t}\int_A u_i(x,y,z,t_k) dA\,, 
\end{equation}
for $i=x,y$ and $N_t$ the number of snapshots. For all runs, $N_t=20$ to obtain equidistant and  statistically independent realizations of the flow. Figure \ref{fig:long_run_mean} displays the result of this analysis for $Ra=10^9$. In panel (a), the mean profile of the $x$-velocity component is shown as a function of even longer averaging times, which were varied from $\tau_{\rm total}=400\, T_f$ (= 20 snapshots) to $1600\, T_f$ (= 80 snapshots). The profile converges steadily to zero, though not uniformly. There is essentially no mean flow. If we insist upon fitting the near wall mean profiles to the two-dimensional Blasius solution, for example, the result is shown in panel (b). In the absence of a definable leading edge distance $x$, we match $\langle u_x\rangle_{A,t}(z)/U_{\infty}$ and $z/\delta_{\infty}$ to $f^{\prime}(\eta)$ which is reported in panels (b,c) of figure \ref{fig:long_run_mean}. Recall that the Blasius solution $f(\eta)=\psi(x,z)/\sqrt{x U_{\infty} \nu}$ and $\eta=z/\delta(x)$ \citep{Schlichting:book:BL}, where $\psi(x,z)$ is the stream function and $u_x=\partial \psi/\partial z$ and $u_z=-\partial \psi/\partial x$. The numerical profiles are rescaled such that the first local maximum of $\langle u_x\rangle_{A,t}$ corresponds to $U_{\infty}$. The fits are shown for different time intervals in (b). Recall that at distance $\eta=5$ the Blasius profile reaches a streamwise velocity magnitude of $f^{\prime}(\eta)=0.99 U_{\infty}$. In this case, panel (a) shows that the maximum velocity reached is of the order 1\% or less of the free-fall velocity, which is the only characteristic velocity of the flow.  As shown in panel (c), no clear trend of the velocity $U^{x,y}_{\infty}$ with averaging time is detectable, and the magnitude is between ${\cal O}(10^{-3})$ and ${\cal O}(10^{-2})$. In table \ref{tab:bl_fit_ux}, the results for all Rayleigh numbers and both horizontal components are listed.

The boundary layer thickness parameters vary when $x$- and $y$-directions are compared at fixed $Ra$. They decrease with increasing Rayleigh number. Furthermore, we calculate the corresponding shear Reynolds numbers $Re_{\rm shear}=U_{\infty}\delta_{\infty}/\nu$, which are found to be very small for all cases. Panels (d) and (e) of figure \ref{fig:long_run_mean} show the mean profiles for all 7 simulation runs and for both horizontal velocity components. They underline the very small mean-flow amplitudes for all Rayleigh numbers in this series. Panel (f) of the figure compares three runs at $Ra=10^9$, which are indicated with a dagger symbol in table \ref{tab:sim_details}, at aspect ratios $\Gamma=2$, 4, and 8. Again, the mean flow amplitudes $\langle u_x\rangle_{A,t}(z)$ are comparable and very small such that an aspect ratio dependence for this result can be excluded when periodic boundary conditions in the horizontal directions are applied. The non-monotonic behavior is simply a reflection of the long averaging times required for convergence in convection studies.

\section{Fluctuation profiles of temperature and velocity}
\begin{figure}
\centerline{\includegraphics[width=0.95\textwidth]{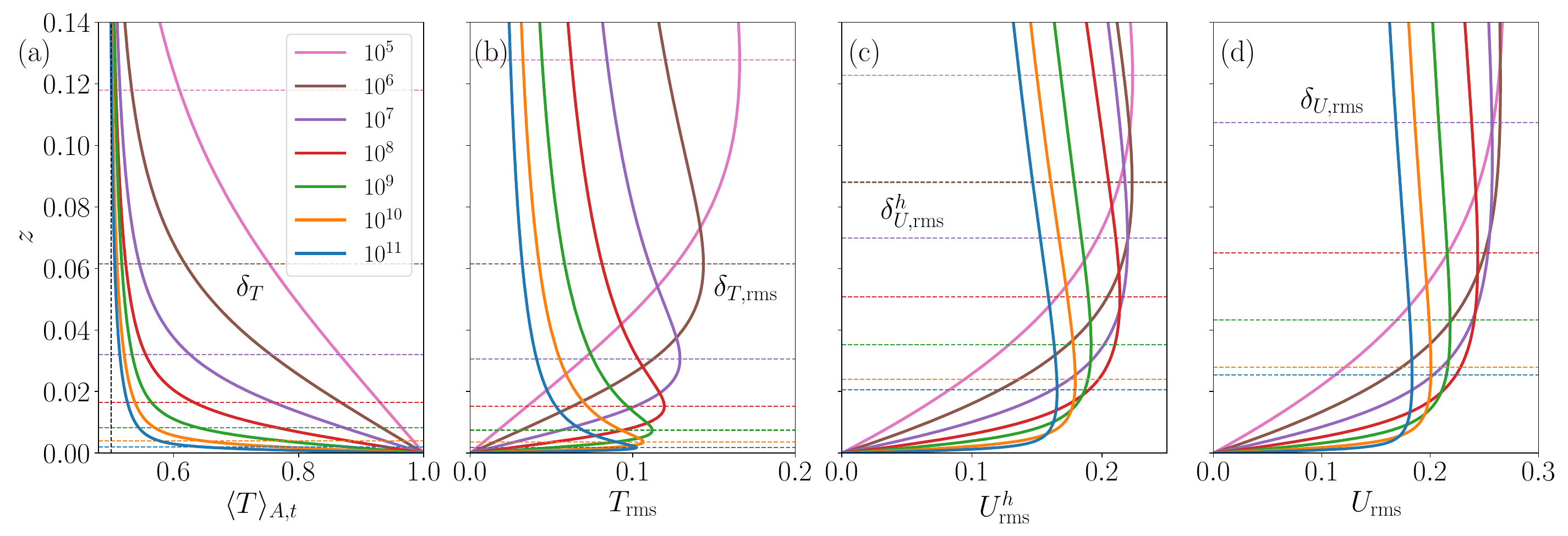}}
  \caption{Mean profiles are compared to each other for all 7 runs. We show the mean profile of temperature $\langle T\rangle_{A,t}(z)$ in panel (a), root-mean-square profile of temperature $T_{\rm rms}(z)$ in (b), root-mean-square profile with respect to the two horizontal velocity components, $U^h_{\rm rms}(z)$, in (c), and with respect to all three components, $U_{\rm rms}(z)$, in (d). The corresponding height scales of these profiles are indicated by dashed horizontal lines in the figures and listed in table~\ref{tab:bl_heights}. Note that for the two lowest $Ra$, they exceed the displayed plot range of the figure in panel (d).}
\label{fig:bl_profiles_compare}
\end{figure}

The mean vertical profiles of the root-mean-square velocities are given by  
\begin{equation}
U_{\rm rms}^h(z)=\sqrt{\langle u_x^2+u_y^2\rangle_{A,t}}\quad\mbox{and}\quad
U_{\rm rms}(z)=\sqrt{\langle u_x^2+u_y^2+u_z^2\rangle_{A,t}}\,,
\label{profile}
\end{equation}
where we distinguish between horizontal and full profiles. The fluctuation profiles for the velocity fields are obtained from the full components since the means are so small. We have verified that the differences in the procedure produce only very small changes.
\begin{table}
  \begin{center}
    \begin{tabular}{lcccccc}
    $Ra$      & $U^x_\infty$            & $U^y_\infty$            & $\delta^x_\infty$     & $\delta^y_\infty$      & $Re^x_{\rm shear}$ & $Re^y_{\rm shear}$ \\
    $10^{ 5}$ & $4.8 \times 10^{-4}$ & $1.7 \times 10^{-3}$ & $9.1 \times 10^{-1}$ & $6.5 \times 10^{-1}$  & $0.16$        & $0.43$        \\
    $10^{ 6}$ & $2.5 \times 10^{-3}$ & $8.4 \times 10^{-3}$ & $1.7 \times 10^{-1}$ & $1.5 \times 10^{-1}$  & $0.51$        & $1.5$        \\
    $10^{ 7}$ & $3.0 \times 10^{-3}$ & $2.7 \times 10^{-3}$ & $8.2 \times 10^{-2}$ & $8.1 \times 10^{-2}$  & $0.93$        & $0.82$        \\
    $10^{ 8}$ & $1.2 \times 10^{-3}$ & $6.6 \times 10^{-3}$ & $5.8 \times 10^{-2}$ & $3.7 \times 10^{-2}$  & $0.80$        & $2.9$        \\
    $10^{ 9}$ & $2.5 \times 10^{-4}$ & $9.3 \times 10^{-4}$ & $5.1 \times 10^{-2}$ & $2.7 \times 10^{-2}$  & $0.48$        & $0.94$        \\
    $10^{10}$ & $3.3 \times 10^{-3}$ & $1.8 \times 10^{-3}$ & $1.3 \times 10^{-2}$ & $1.1 \times 10^{-2}$  & $5.1$        & $2.4$        \\
    $10^{11}$ & $3.6 \times 10^{-3}$ & $1.4 \times 10^{-4}$ & $5.8 \times 10^{-3}$ & $7.0 \times 10^{-3}$  & $7.9$        & $0.38$        \\
    \end{tabular}
    \caption{Fit parameters $U^x_{\infty}$ and $U^y_{\infty}$ as well as the corresponding thicknesses $\delta^x_{\infty}$ and $\delta^y_{\infty}$ of the combined plane-time averaged mean profiles of the horizontal velocity components $\langle u_x\rangle _{A,t}$ and $\langle u_y\rangle_{A,t}$ to the Blasius profile $U_{\infty}f^{\prime}(\eta)$ with $\eta=z/\delta_{\infty}$. The corresponding shear Reynolds numbers $\Rey^{x,y}_{\mathrm{shear}}=U^{x,y}_{\infty}\delta^{x,y}_{\infty}/\nu$ are also given. All runs are at $\Gamma=4$.}
    \label{tab:bl_fit_ux}
  \end{center}
\end{table}
Figure \ref{fig:bl_profiles_compare} summarizes mean profiles for 7 simulation runs, the mean temperature profile and the root-mean-square profiles of temperature, horizontal velocity components, and all three velocity components. The temperature fluctuation profile is similarly obtained by $T_{\rm rms}(z)=\langle\theta^2\rangle^{1/2}_{A,t}$ with $\theta({\bm x},t)=T({\bm x},t)-\langle T\rangle_{A,t}(z)$. The correspondingly related characteristic scales are indicated by horizontal dashed lines and detailed in table \ref{tab:bl_heights}. It is seen that the thermal boundary layer thickness $\delta_T=1/(2 Nu)$ is slightly larger than the distance from the wall of the maximum of the temperature fluctuation profile, which we term the thermal fluctuation boundary layer thickness or, for short, {\em thermal fluctuation thickness}, $\delta_{T,{\rm rms}}$. Increasingly larger are distances from the wall to the maxima of the velocity fluctuation profiles, obeying a ratio of $\delta_{U,{\rm rms}}/\delta_{T,{\rm rms}}\approx 2$ for $Ra=10^5$ up to approximately 13 for $Ra=10^{11}$; see panel (d). The corresponding thicknesses are termed {\em velocity fluctuation thickness}.

\begin{figure}
\centerline{\includegraphics[width=0.95\textwidth]{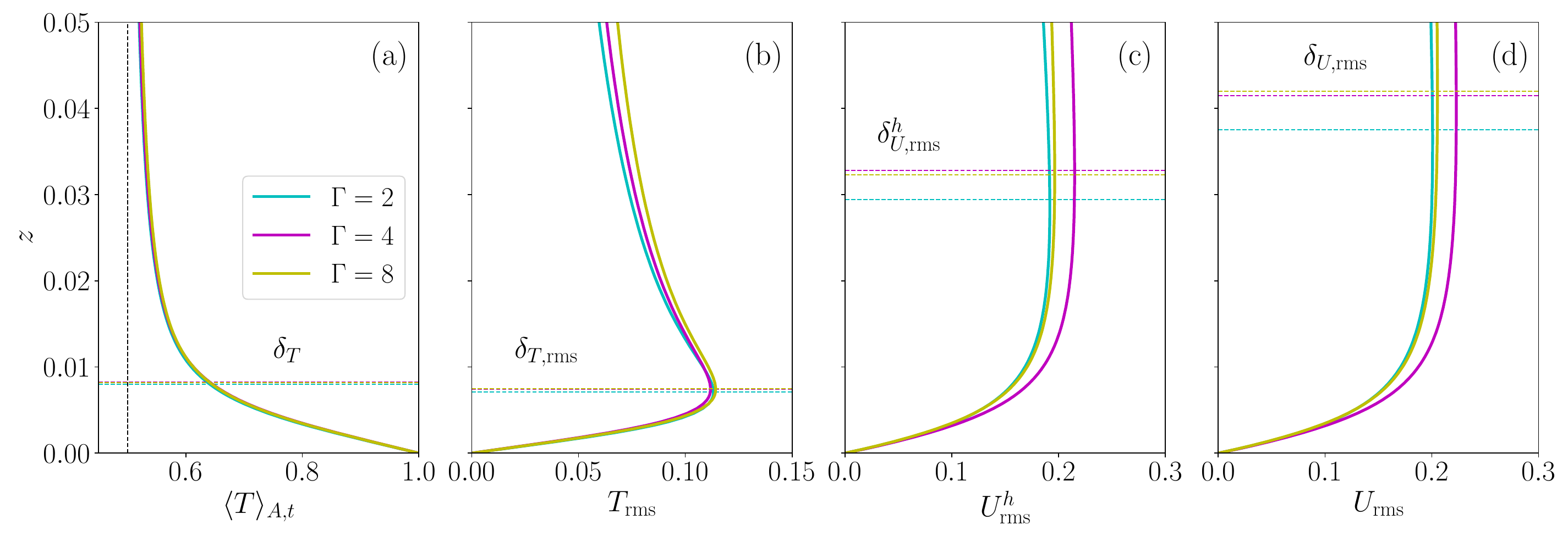}}
  \caption{Mean temperature profiles compared to each other for different aspect ratio at $Ra=10^9$. Similar to figure \ref{fig:bl_profiles_compare}, we show the mean profile of temperature $\langle T\rangle_{A,t}(z)$ in panel (a), root-mean-square profile of temperature $T_{\rm rms}(z)$ in (b), root-mean-square profile with respect to the horizontal velocity components, $U^h_{\rm rms}(z)$ in (c), and with respect to all three components $U_{\rm rms}(z)$ in (d). The corresponding height scales of these profiles are indicated by dashed horizontal lines and are also listed in table~\ref{tab:bl_heights}.}
\label{fig:gamma}
\end{figure}

Furthermore, we repeated the fluctuation profile analysis for the dependence on different aspect ratios, one on each side of 4, at $Ra=10^9$. These runs are indicated again by a dagger symbol in table \ref{tab:bl_heights}, where we collect the corresponding thickness scales. The corresponding profiles are shown in figure \ref{fig:gamma}. While the temperature profiles collapse close to the walls, thus displaying no sensitivity with respect to the aspect ratio $\Gamma$ in this range, the velocity profiles are affected by $\Gamma$. However, the resulting velocity fluctuation thicknesses are found to agree well for $\Gamma=4$ and 8 (by 1.5\% or less), in terms of both horizontal and full velocity fluctuations. The finding supports our considered view that $\Gamma\ge 4$ is sufficient to obtain horizontal homogeneity for the statistics already introduced.
\begin{table}
  \begin{center}
    \begin{tabular}{lccccc}
    $Ra$         & $\Gamma$ & $\delta_T$                        & $\delta_{T,{\rm rms}}$            & $\delta^h_{U,{\rm rms}}$          & $\delta_{U,{\rm rms}}$            \\
    $10^{ 5}$           & 4 & $(1.17 \pm 0.04) \times 10^{-1}$  & $(1.27 \pm 0.08) \times 10^{-1}$  & $(1.23 \pm 0.07) \times 10^{-1}$  & $(2.39 \pm 1.32) \times 10^{-1}$  \\
    $10^{ 6}$           & 4 & $(6.13 \pm 0.15) \times 10^{-2}$  & $(6.13 \pm 0.27) \times 10^{-2}$  & $(8.80 \pm 0.54) \times 10^{-2}$  & $(1.74 \pm 1.12) \times 10^{-1}$  \\
    $10^{ 7}$           & 4 & $(3.21 \pm 0.06) \times 10^{-2}$  & $(3.07 \pm 0.10) \times 10^{-2}$  & $(6.99 \pm 0.58) \times 10^{-2}$  & $(1.07 \pm 0.51) \times 10^{-1}$  \\
    $10^{ 8}$           & 4 & $(1.64 \pm 0.02) \times 10^{-2}$  & $(1.52 \pm 0.04) \times 10^{-2}$  & $(5.08 \pm 0.35) \times 10^{-2}$  & $(6.51 \pm 0.92) \times 10^{-2}$  \\
    $10^{ 9}$           & 4 & $(8.22 \pm 0.12) \times 10^{-3}$  & $(7.47 \pm 0.18) \times 10^{-3}$  & $(3.52 \pm 0.28) \times 10^{-2}$  & $(4.33 \pm 1.97) \times 10^{-2}$  \\
    $10^{ 9\,\dagger}$  & 2 & $(8.01 \pm 0.15) \times 10^{-3}$  & $(7.07 \pm 0.29) \times 10^{-3}$  & $(2.94 \pm 0.49) \times 10^{-2}$  & $(3.75 \pm 0.72) \times 10^{-2}$  \\
    $10^{ 9\,\dagger}$  & 4 & $(8.27 \pm 0.10) \times 10^{-3}$  & $(7.44 \pm 0.21) \times 10^{-3}$  & $(3.28 \pm 0.28) \times 10^{-2}$  & $(4.15 \pm 0.63) \times 10^{-2}$  \\
    $10^{ 9\,\dagger}$  & 8 & $(8.20 \pm 0.04) \times 10^{-3}$  & $(7.51 \pm 0.12) \times 10^{-3}$  & $(3.23 \pm 0.12) \times 10^{-2}$  & $(4.20 \pm 0.22) \times 10^{-2}$  \\
    $10^{10}$           & 4 & $(4.06 \pm 0.04) \times 10^{-3}$  & $(3.67 \pm 0.11) \times 10^{-3}$  & $(2.39 \pm 0.19) \times 10^{-2}$  & $(2.79 \pm 0.27) \times 10^{-2}$  \\
    $10^{11}$           & 4 & $(1.97 \pm 0.02) \times 10^{-3}$  & $(1.81 \pm 0.05) \times 10^{-3}$  & $(2.07 \pm 0.27) \times 10^{-2}$  & $(2.55 \pm 0.39) \times 10^{-2}$  \\
    \end{tabular}
    \caption{
    Differently defined boundary layer thicknesses of temperature and velocity according to the vertical mean profiles plotted in figure \ref{fig:bl_profiles_compare}. The standard deviation accompanies each mean value.}
    \label{tab:bl_heights}
  \end{center}
\end{table}


\section{Scaling of combined volume-time averaged fluctuation with Rayleigh number}

\begin{figure}
\centering
\includegraphics[width=0.8\textwidth]{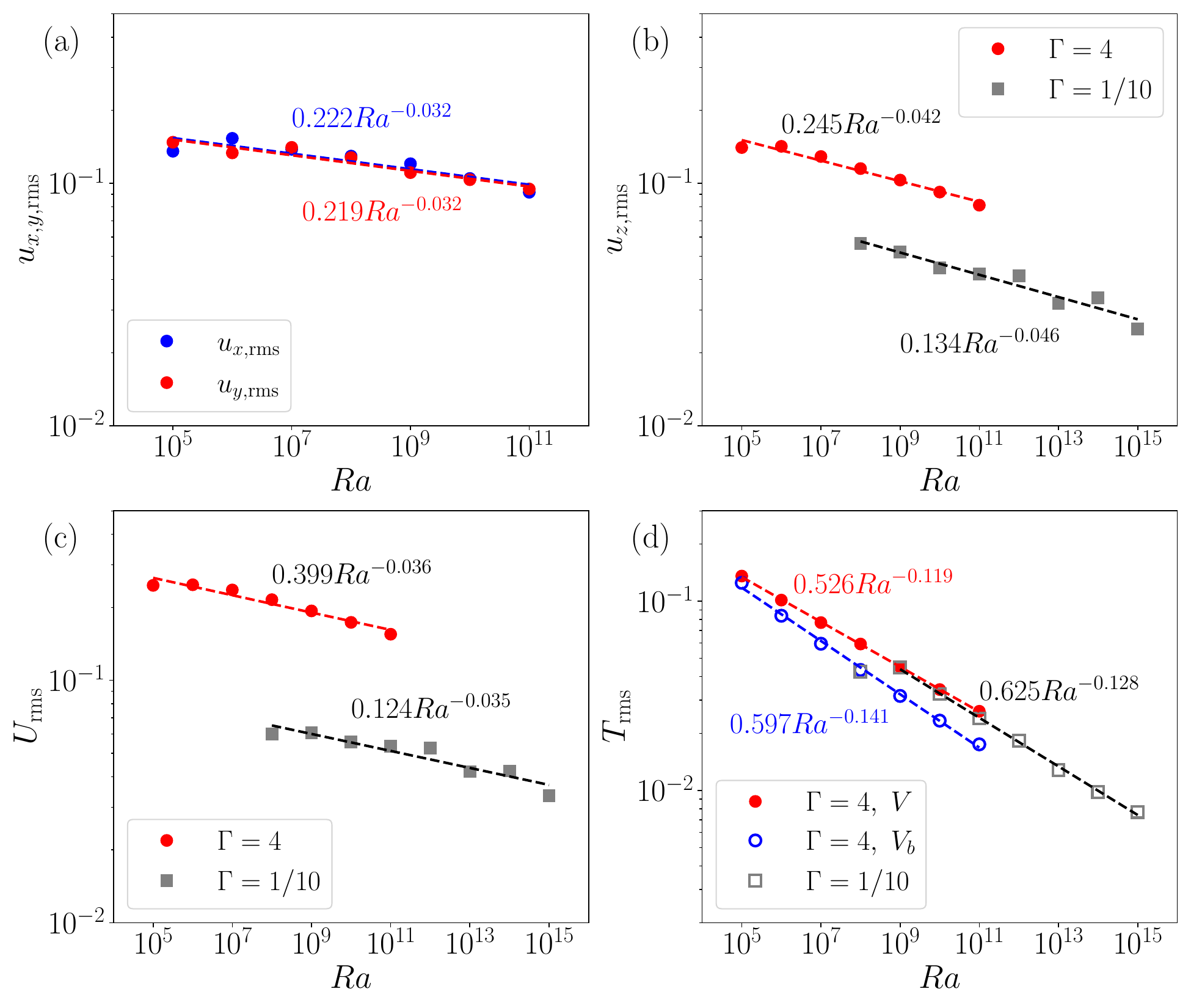}
  \caption{Scaling of the root-mean-square velocities and temperature with respect to Rayleigh number. All values are obtained by a combined average with respect to volume and time. (a) Horizontal velocity components. (b) Vertical velocity component. (c) All three velocity components. (d) Temperature. Dashed lines denote power law fits to the data. Blue open circles stand for data obtained in the bulk volume $V_b$. In panels (b--d), we add data from \cite{Iyer:PNAS2020} with separate fits. The corresponding temperature data are obtained there for the cylindrical cell interior, $r\le 0.03 H$ and $Ra\ge 10^9$.}
\label{fig:fluctuations}
\end{figure}
Figure \ref{fig:fluctuations} summarizes the root-mean-square fluctuations of the three velocity components and the temperature. They are obtained by a combined average with respect to the full volume $V=L^2H$ and time, e.g. $u_{x,{\rm rms}}=\langle u_x^2\rangle^{1/2}_{V,t}$. The quantity $U_{\rm rms}$ denotes again the fluctuations with respect to all three velocity components. It is seen that the dependence of the velocity fluctuations on the Rayleigh number is very weak with $\beta \lesssim 0.042$. The temperature fluctuations drop with a smaller power law exponent, $T_{\rm rms}\sim Ra^{-\beta}$, which is found to be $\beta=0.119$ for the present data. This exponent is slightly smaller in magnitude than those reported in experiments in cylindrical cells of aspect ratio 1/2. For comparison, \cite{Castaing:JFM1989}, \cite{Niemela:Nature2000} and \cite{Wu1992} report exponents of $\beta\approx 0.145$. We also analysed the temperature fluctuations in the bulk of the layer, which takes a volume average with respect to $V_b=L^2\times [0.4,0.6]$ and time. The exponent changes to $\beta=0.141$ which is closer to the experiments. We have verified that a variation of the thickness of the bulk volume $V_b$ does not alter the results significantly. 

In panels (b--d) of figure \ref{fig:fluctuations}, we added data from the DNS of \cite{Iyer:PNAS2020} for comparison, which were obtained in a closed slender cylindrical cell of aspect ratio $\Gamma=0.1$. It is seen that exponents of the power law fits are close to those of the present simulation series. The prefactors differ as expected, because the former DNS data were obtained for geometrically constrained convection. This finding of nearly the same scaling exponents suggests a robust geometry-independent trend of all thermal fluctuations with respect to the Rayleigh number. Geometry-specific aspects mostly affect the prefactor. 

\section{Decomposition into coherent and incoherent boundary layer regions}
\begin{figure}
\centering
\includegraphics[width=0.99\textwidth]{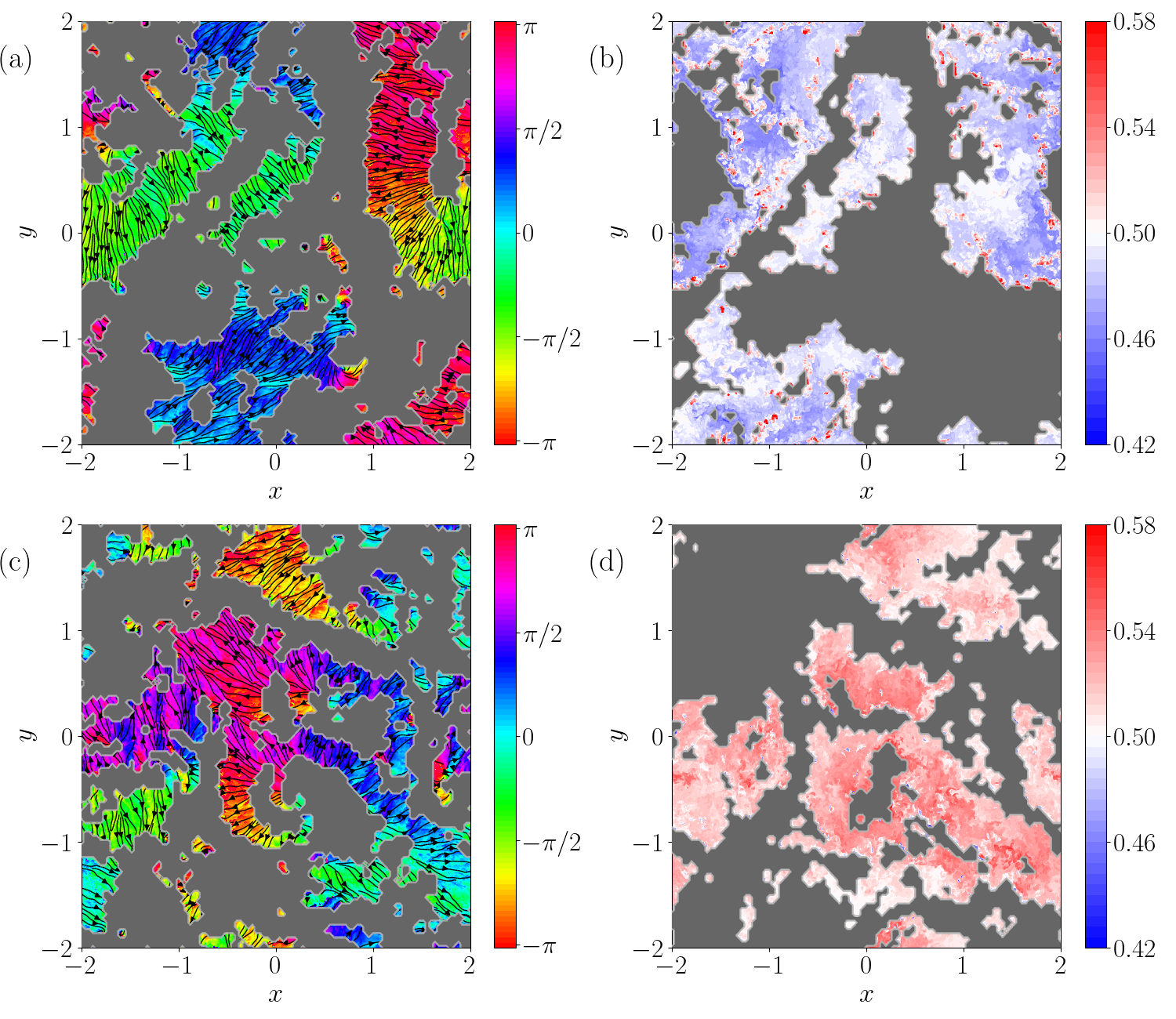}
  \caption{Boundary layer flow structure for $Ra=10^{10}$ at the top and bottom walls. (a) Decomposition of a snapshot into coherent and incoherent (in gray) flow regions at $z=\delta_T$. For the coherent shear-dominated regions, we overlay horizontal streamlines and indicate their local flow direction. (b) Corresponding temperature field in plane at $z=0.1\approx 25 \delta_T$. Gray area corresponds to $T>T_0=0.5$. (c) Same snapshot with the decomposition similar to (b) for $z=H-\delta_T$ close to the top wall. (d) Corresponding temperature field at $z=0.9\approx 25 \delta_T$. Gray area corresponds now to $T<T_0$. Note that all plots are shown at the coarse resolution of $100\times 100$ square boxes (which were used to calculate the local mean flow magnitude and orientation) and not at the original spectral resolution of the DNS run.}
\label{fig:coh}
\end{figure}

The orientation of the boundary layer flow varies strongly as shown in figures \ref{fig:coh} (a) and (c), where we plot the orientation angle of the horizontal velocity $\varphi=\arctan(u_y/u_x) \in [-\pi,\pi]$ for a snapshot at $Ra=10^{10}$ at $z=\delta_T$ and $z=H-\delta_T$, respectively. At both these heights, we cover the horizontal plane into $10^4$ disjoint square boxes of area content $A_i=A/10^4$, where $A=L^2$.  We then calculate the mean horizontal velocity $\bar{\bm u}_h(A_i)$ in each of the $A_i$ and decompose the cross section into {\em coherent} and {\em incoherent} boundary layer regions for $|\bar{\bm u}_h(A_i)| > U^h_{\rm rms}(\delta_T)$ and $|\bar{\bm u}_h(A_i)|\le U^h_{\rm rms}(\delta_T)$, respectively. Similar decompositions have been applied to analyse the spatio-temporal intermittency of the transition to turbulence in shear flow turbulence in extended domains, see e.g. \citet{Hof2022}. Panels (a, c) of figure \ref{fig:coh} show that coherent shear-dominated patches are separated by incoherent flow regions (in gray). See also figure \ref{fig:contours}. The superposed streamlines indicate the different flow orientations of the shear-dominated regions.

Panels (b,d) of the same figure show the corresponding snapshots of the temperature field $T$ at $z=0.1$ above the bottom and $z=0.9$ below the top, which are distances of 25 $\delta_T$ away from the walls at $Ra=10^{10}$. It is clearly seen that the hotter regions at $z=0.1$ and the colder regions at $z=0.9$, both of which are displayed in gray, coincide fairly well with the incoherent flow regions at the edge of the thermal boundary layer. We can define overlap factors $0\le \tilde{O}\le 1$ by
\begin{equation}
\tilde{O}_{\rm bot}=\frac{A_{\rm incoh} \bigcup A_{T>T_0}}{\max(A_{\rm incoh},A_{T>T_0})} 
\quad\mbox{and}\quad
\tilde{O}_{\rm top}=\frac{A_{\rm incoh} \bigcup A_{T<T_0}}{\max(A_{\rm incoh},A_{T<T_0})}\,, 
\end{equation}
with $T_0=0.5$. Here, we find mean overlaps of $\langle\tilde{O}_{\rm bot}\rangle=0.60$ and $\langle\tilde{O}_{\rm top}\rangle=0.63$, where the average is taken over the snapshots. The physical interpretation is as follows: the incoherent regions correspond to dominant hotter upwelling (colder downwelling) motions. These regions occur outside shear-dominated patches where the thermal plumes merge successively with growing distances from the walls. As one approaches the mid-plane of the convection cell, they tend to form the turbulent superstructure pattern of convection. We have determined that the area fraction of the incoherent regions remains nearly constant at approximately 60\% of $A$ for the whole Rayleigh number range. The insensitivity of the volume fractions with respect to the Rayleigh number suggests that this skeleton of upwelling (downwelling) incoherent regions could be a relic from the weakly nonlinear regime of convection at much lower Rayleigh numbers, which itself arises from the onset of convection by a linear primary instability, filling the whole domain with convection rolls. 

We have varied the threshold for this analysis from $0.5 U^h_{\rm rms}(\delta_T)$ to $2 U^h_{\rm rms}(\delta_T)$. While the incoherent fractions do depend on the threshold when its variations are large, they are practically independent of the Rayleigh number even for such large variations stated above. This supports our choice of $U^h_{\rm rms}(\delta_T)$ as a physically meaningful threshold.

\begin{figure}
\centering
\includegraphics[width=0.99\textwidth]{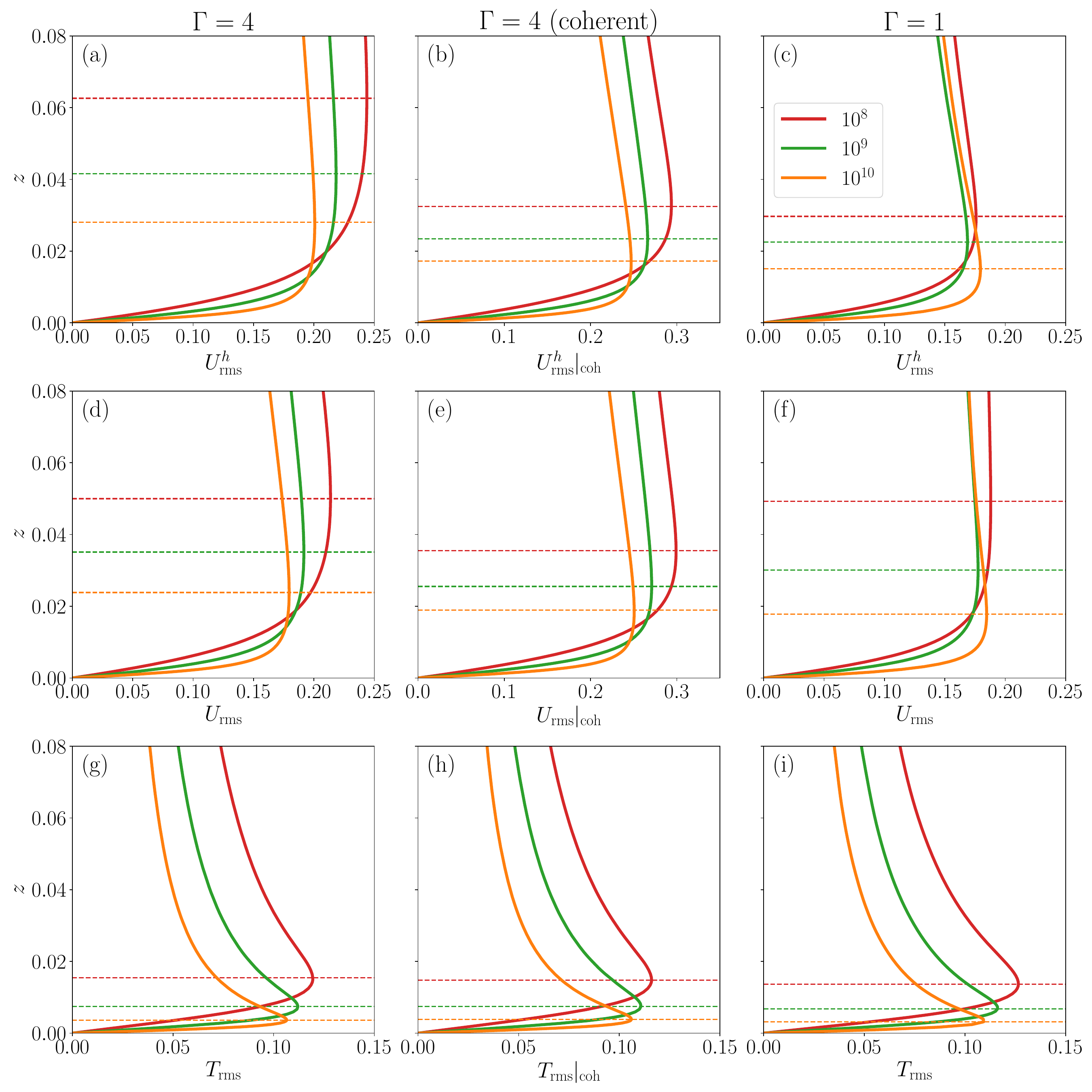}
  \caption{Vertical profiles of the velocity and temperature fluctuations for Rayleigh numbers $Ra=10^8$, $10^9$, and $10^{10}$. The figure compares averages over the full cross section at $\Gamma=4$ (left column) with averages over the coherent shear-dominated regions (middle column) and with averages over the full circular cross section in closed cylindrical cell at  $\Gamma=1$ (right column) from \citet{Scheel:PRF2017}. Panels (a,b,c) are for $U^h_{\rm rms}$, panels (d,e,f) for $U_{\rm rms}$, and panels (g,h,i) for $T_{\rm rms}$. The dashed lines in each panel mark the location of the first local maximum away from the wall, and indicate the corresponding fluctuation thickness. The color coding, which is indicated in panel (c), holds for all panels.}
\label{fig:cond}
\end{figure}

We can now return to the fluctuation analysis which is conditioned on coherent and incoherent regions in the following. Figure \ref{fig:cond} replots the rms profiles of full and horizontal velocity and temperature profiles for Rayleigh numbers $Ra=10^8$, $10^9$, and $10^{10}$. We have chosen these three Rayleigh numbers of our series to provide a one-to-one comparison with DNS data in a closed cylindrical cell at $\Gamma=1$ of \cite{Scheel:PRF2017}. They are also shown in the figure. Vertical profiles, which have been taken over the full cross section (denoted as case G4 in the following), are shown in the left column of figure \ref{fig:cond}. Profiles conditioned on shear-dominated regions are displayed in the middle column (case G4C), while those for the cylindrical cross section of the closed container (case G1) are shown in the right column. From the bottom row of the figure, it is clear that the temperature profiles of G4, G4C, and G1 for all three $Ra$ agree. This suggests that the temperature boundary layers are alike in all cases. This is different for the velocity field, for which the horizontal velocity fluctuations (displayed in top row) show a clear trend. The thickness scale decreases from G4 to G4C and even more from G4C to G1. The close agreement of G4C and G1 clearly supports the dominance of shearing motion in the boundary regions in closed cylindrical cells, imposed by the prominent large-scale circulation. It is in line with a reduced fluctuation thickness. For fluctuations with respect to the full velocity field, we do detect a decrease of the thickness from G4 to G4C, but not from G4C to G1 for the two lower $Ra$. We suspect that this might be caused by prominent coherent up- and downwelling motions at the side walls for the lower $Ra$ which effectively enhance the thickness \citep{Schumacher2016}.

Unlike G1 and G4C, for which the velocity and temperature boundary layers have comparable thicknesses, the G4 case shows that the velocity boundary layer is much thicker than the temperature thickness, suggesting a different mechanism in G4. We recall that the notion of the velocity boundary layer is only nominal in the sense that they are based on fluctuation profiles and the mean velocity variation within that region is quite small (see figure 4).

\begin{figure}
\centering
\includegraphics[width=0.6\textwidth]{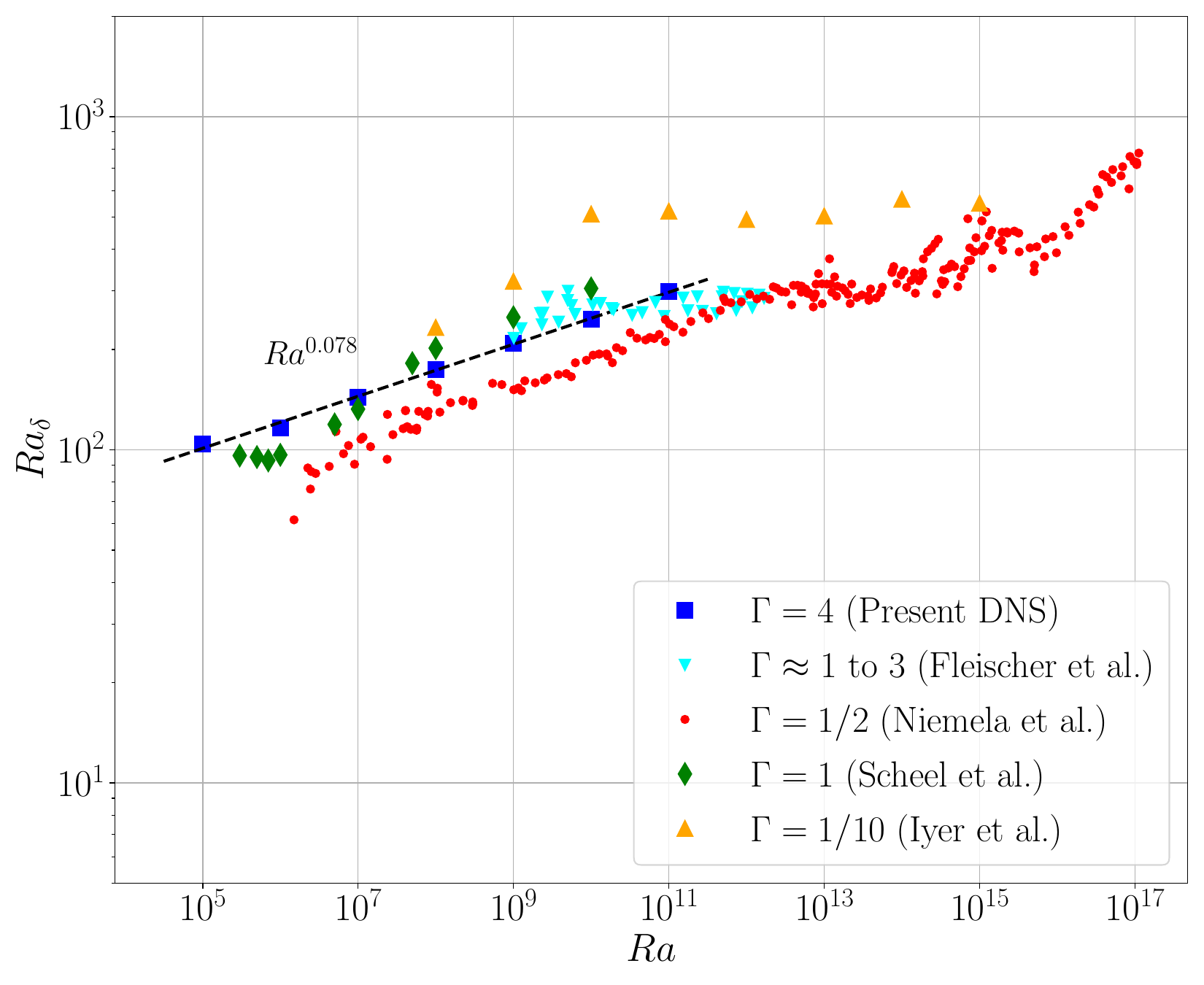}
  \caption{Boundary layer thickness-based Rayleigh number $Ra_{\delta}$ versus Rayleigh number $Ra$. We have collected for comparison data from other simulations including \citet{Scheel:PRF2017} and \citet{Iyer:PNAS2020}; also plotted are experimental data from \citet{Fleischer:JFM2002} and \citet{Niemela:Nature2000}. Note that all listed data from previous sources use $\delta_T=1/(2 Nu)$ to calculate the boundary layer thickness-based Rayleigh number $Ra_{\delta}$. Only the present data use $\delta_{T, {\rm rms}}$ for the determination of $Ra_{\delta}$. The corresponding Rayleigh numbers $Ra_{\delta}=Ra_{\delta,{\rm rms}}$ are found in table \ref{tab:sim_details}.}
\label{fig:radelta}
\end{figure}

\section{Final discussion}
Our DNS of the turbulent Rayleigh-B\'{e}nard convection encompasses a Cartesian domain with $\Gamma=4$, with no-slip horizontal walls and periodic boundary conditions for the side faces. These simulations up to $Ra=10^{11}$ are aimed to approach the original canonical case of a plane convection layer between a pair of infinitely extended rigid plates. We demonstrated that a standard mean flow profile (obtained by combining averages with respect to time and the entire horizontal cross sectional plane) have very small magnitudes, and that efforts to match them to laminar boundary layer profiles produced no conclusive results. To the extent that we can define the boundary layers, they give very small shear Reynolds numbers (see below). In the long-time limit, which we have followed for 1600 $T_f$ at $Ra=10^9$, the velocity mean profiles have to converge to $\mbox{lim}_{t\to \infty}\langle u_i\rangle_{A,t}(z)\to 0$ due to statistical homogeneity in $x$ and $y$. The simulations by \cite{Hartlep:PRL2003} (for $Ra\leq 1\times 10^7$) also showed that the mean flow contained very little kinetic energy, but  \cite{De:EJMB2018} found a long time periodicity in the mean flow for low Rayleigh numbers, $Ra\leq 2\times 10^6$.

Rather than having a mean flow profile with small velocity fluctuations, we are faced with small mean velocity amplitudes in the presence of velocity fluctuations that are up to 2 orders of magnitude larger when the statistics are taken over finite time intervals $\tau_{\rm total}$, as seen from comparisons of table \ref{tab:bl_fit_ux} with the data in figure \ref{fig:bl_profiles_compare}. This central result also holds when the aspect ratio of the simulation is varied. It is our view that fluctuations will be relevant for all configurations which includes closed cells of $\Gamma\lesssim 1$, see e.g. figure \ref{fig:cond}. But their relevance is strongest in the statistically homogeneous plane layer with periodic boundary conditions in the horizontal direction---the configuration that comes closest to the original physical problem of turbulent convection \citep{Spiegel1962}, as relevant for most geo- and astrophysical applications.

We also showed that the corresponding shear Reynolds numbers, which are based on mean flow quantities for a finite averaging time, remain very small because the characteristic velocities $U_{\infty}$ are small. The strong fluctuations cause the fluctuation thicknesses of temperature and velocity (defined as the near-wall maxima of the rms profiles) to differ by an order of magnitude for the highest Rayleigh numbers, as summarized in table \ref{tab:bl_heights}. This difference increases with Rayleigh number (though the Prandtl number is held fixed at order unity); it becomes particularly pronounced for $Ra\ge 10^9$, a range beyond which previous larger-aspect-ratio DNS studies rarely advanced. 

Furthermore, our analysis revealed that the velocity boundary region in the present configuration is a carpet of differently oriented time-dependent and shear-dominated (coherent) regions interspersed by regions of incoherent flow. The latter regions occupy about 60\% of the plate area for all Rayleigh numbers. This heterogeneous composition crystallizes particularly for $Ra\gtrsim 10^9$, underlying again the importance of DNS with larger aspect ratios {\em and} high Rayleigh numbers. The incoherent regions in the present flow can be as large as the entire cross section of a cylindrical cell at $\Gamma\sim 1$. The coherent regions are the near-wall footprint of the circulation rolls which form the large-scale turbulent superstructure pattern \citep{Stevens:PRF2018,Pandey:NC2018}. They change their orientation continually and thus result in a net zero mean flow as stated above. 

Finally, we showed that the velocity fluctuation thicknesses decrease when they are conditioned on shear-dominated patches in the near-wall region. They are then closer to those scales which are obtained in turbulent convection in closed cylindrical cells of aspect ratio $\Gamma\lesssim 1$. The geometry of the closed cell enforces a large-scale circulation which is mostly shear-dominated in the vicinity of the walls, as already shown in \citet{Schumacher:PRF2016} where the time-dependence of the orientation has been eliminated. This causes smaller velocity thickness scales that are, however, still larger than the thermal boundary layer thickness at $Pr\sim 1$. Nevertheless, the fluctuation thickness is the consistently definable velocity boundary layer scale for the present turbulent convection flow.

The present results also raise many questions on the possible transition mechanisms of the boundary layer to a turbulent regime and the possible consequences for the global heat transfer. Differently from wall-bounded shear flows, we detect velocity fluctuations everywhere, even though at different strengths; see again figure \ref{fig:cond}. Furthermore, we do not observe a Rayleigh number dependence of the ratio of coherent (``laminar") to incoherent (``turbulent") regions. The time scales, at which these complex spatio-temporal patterns change, become increasingly shorter with increasing Rayleigh number. A variation of the threshold for the decomposition into coherent and incoherent boundary regions practically did not affect this Rayleigh number independence.

The spatio-temporal heterogeneity of the velocity boundary layer, which we detected here, suggests to us the prevalence of local, rather than global, instability mechanisms, which would bring us back to the marginal stability concept of the boundary layer, see e.g. \citet{Howard1966} and for a detailed boundary layer model with plume formation, \citet{Theerthan1998}. However, figure \ref{fig:radelta} shows a power law fit of $Ra_{\delta,{\rm rms}}=A Ra^\gamma$ with a very small exponent $\gamma=0.077$ and $A\approx 41$. The resulting $Ra_{\delta,{\rm rms}}$ are by at least a factor of 4 smaller than Howard's critical Rayleigh number of $Ra_{\delta}\sim {\cal O}(10^3)$. Additional data from experiments show that even at $Ra\sim 10^{17}$ a thickness-based Rayleigh number $Ra_{\delta}$ barely reaches a value of $10^3$. This challenges the original marginal stability concept. This question and higher Rayleigh number simulation in the present configuration form the subject of further study.

\section*{Acknowledgements}
The authors thank Jaywant Arakeri, Björn Hof, Hassan Nagib and Peter Schmid for helpful discussions.

\section*{Funding}
The work of RJS is funded by the European Union (ERC, MesoComp, 101052786). Views and opinions expressed are however those of the authors only and do not necessarily reflect those of the European Union or the European Research Council. The work of JDS was supported by a Mercator Fellowship of the Deutsche Forschungsgemeinschaft within the Priority Programme DFG-SPP 1881 on Turbulent Superstructures. The authors gratefully acknowledge the Gauss Centre for Supercomputing e.V. (https://www.gauss-centre.eu) for funding the project nonbou by providing computing time through the John von Neumann Institute for Computing (NIC) on the GCS Supercomputer JUWELS Booster at Jülich Supercomputing Centre (JSC). 

\section*{Declaration of interests}
The authors report no conflict of interest.

\section*{Data availability statement}
The data that support the findings of this study are available on reasonable request.

\section*{Author ORCIDs}
Roshan J. Samuel https://orcid.org/0000-0002-1280-9881

\noindent
Mathis Bode https://orcid.org/0000-0001-9922-9742

\noindent
Janet D. Scheel https://orcid.org/0000-0002-1669-4188

\noindent
Katepalli R. Sreenivasan https://orcid.org/0000-0002-3943-6827

\noindent
J{\"o}rg Schumacher https://orcid.org/0000-0002-1359-4536

\section*{Author contributions}
All authors designed the research. R.J.S., M.B. and J.S. carried out the supercomputer simulations. R.J.S., J.D.S and J.S.  analysed the simulation data and generated the figures. All authors wrote the manuscript. 

\bibliographystyle{jfm}

\end{document}